# Cr$_2$O$_3$ thin films grown at room temperature by low pressure laser chemical vapour deposition

P.M. Sousa[1], A.J. Silvestre[2,a] and O. Conde[1]

[1] *Universidade de Lisboa, Faculdade de Ciências, Departamento de Física and ICEMS, Campo Grande, Ed. C8, 1749-016 Lisboa, Portugal*

[2] *Instituto Superior de Engenharia de Lisboa and ICEMS, R. Conselheiro Emídio Navarro 1, 1959-007 Lisboa, Portugal*

**Abstract**

Chromia (Cr$_2$O$_3$) has been extensively explored for the purpose of developing widespread industrial applications, owing to the convergence of a variety of mechanical, physical and chemical properties in one single oxide material. Various methods have been used for large area synthesis of Cr$_2$O$_3$ films. However, for selective area growth and growth on thermally sensitive materials, laser-assisted chemical vapour deposition (LCVD) can be applied advantageously.

Here we report on the growth of single layers of pure Cr$_2$O$_3$ onto sapphire substrates at room temperature by low pressure photolytic LCVD, using UV laser radiation and Cr(CO)$_6$ as chromium precursor. The feasibility of the LCVD technique to access selective area deposition of chromia thin films is demonstrated. Best results were obtained for a laser fluence of 120 mJ cm$^{-2}$ and a partial pressure ratio of O$_2$ to Cr(CO)$_6$ of 1.0. Samples grown with these experimental parameters are polycrystalline and their microstructure is characterised by a high density of particles whose size follows a lognormal distribution. Deposition rates of 0.1 nm s$^{-1}$ and mean particle sizes of 1.85 µm were measured for these films.

---

[a] Corresponding author: e-mail: asilvestre@deq.isel.ipl.pt

## 1. Introduction

Among the different chromium oxide solid phases, $Cr_2O_3$ is the most stable, existing in a wide range of temperature and pressure [1]. It exhibits many attractive tribological properties, such as high hardness (29.5 GPa) and high melting point (~2300 ºC) combined with chemical inertness, low friction coefficient, high wear resistance, and high temperature oxidation resistance [2,3]. Regarding magnetic behaviour, $Cr_2O_3$ is antiferromagnetic with a Néel temperature of 307 K [4]; however, the antiferromagnetic character can be changed to weak ferromagnetism [5] and even superparamagnetism [6] when chromia nanoparticles are considered. On the other hand, despite its intrinsic insulator nature, $Cr_2O_3$ films can present either *p*-type or *n*-type semiconductor behaviour, depending on the growth conditions [7]. It also shows a high solar absorption coefficient and low thermal emissivity [8]. The confluence of all these properties in a single material makes $Cr_2O_3$ a key material for the development of a broad range of industrial applications.

A wide variety of chemical and physical methods have been used for large area synthesis of $Cr_2O_3$ films, e.g. chemical vapour deposition (CVD) at either atmospheric [9,10] or low pressure [11], plasma enhanced CVD (PECVD) [12,13], electrodeposition [14], metal oxidation [15], chemical spray pyrolysis [16], RF magnetron sputtering [17,18], molecular beam epitaxy [19,20] and atomic layer deposition [21].

Laser-assisted chemical vapour deposition (LCVD) can be an advantageous technique whenever selective area growth or thermally sensitive substrate materials are required. Dowben and co-authors [22] achieved the synthesis of films containing both $Cr_2O_3$ and $CrO_2$ on Si(111) wafers, in static reactive atmosphere at low pressure of ~$10^{-3}$ Pa, by using a nitrogen laser ($\lambda = 337$ nm) and $Cr(CO)_6$ as chromium precursor. Looking for a better efficiency of the deposition process, we have previously explored dynamic atmospheres at higher pressure (~10 Pa) using KrF laser radiation ($\lambda = 248$ nm) for which the $Cr(CO)_6$ absorption cross section ($5\times10^{-17}$ cm$^2$) is much higher than for the nitrogen laser (<$10^{-19}$ cm$^2$) [23,24]. In such experimental conditions biphasic films consisting of $Cr_2O_3$ and $CrO_2$ were also deposited [24].

Here we present results on the synthesis of pure $Cr_2O_3$ films grown onto c-cut sapphire at room temperature by LCVD at low pressure, using KrF laser radiation to promote the photodissociation of $Cr(CO)_6$ in dynamic atmospheres containing oxygen.

## 2. Experimental details

The experimental apparatus used for the depositions consists of a KrF excimer laser (248 nm wavelength, 30 ns pulse duration), a stainless steel high vacuum reaction chamber, a high vacuum pumping system and a gas system for delivering the reactants to the reaction chamber. $Al_2O_3$ (0001) substrates with surface area of 10×5 mm$^2$ and 500 μm thick were used. The KrF laser working at a repetition rate of 10 Hz irradiated the substrates at perpendicular incidence. The beam spot size on the substrate's surface was 8.2 mm$^2$. Prior to any deposition experiment the reactor was evacuated to a base pressure lower than $2\times10^{-4}$ Pa. All the deposition experiments were conducted at room temperature (RT) in the dynamic gas regime, at a constant total pressure of $4.5\times10^{-3}$ Pa and for a total deposition time of 4 hours. $Cr(CO)_6$ powder, whose vapour pressure as a function of temperature is given by $\log_{10} p_{Cr(CO)_6} = 12.75 - 3285/T$ [25], where $p_{Cr(CO)_6}$ is in Pa and $T$ in K, was sublimated at 22 ºC in a controlled temperature stainless steel cell connected to the LCVD reaction chamber. Its flow rate was varied in the range 0.03 - 0.2 sccm by adjusting the pressure difference between the reactor and the cell. The $O_2$ flow rate was varied in the range 0.02 - 0.2 sccm, using a highly sensitive mass flow controller.

The laser fluence, $F$, and the partial pressure ratio of $O_2$ to $Cr(CO)_6$, $R=p_{O2}/p_{Cr(CO)_6}$, were correlated with the phase composition, degree of crystallinity, deposition rate, and microstructure of the as-deposited films. Structural analysis was carried out by X-ray diffraction (XRD) with Cu Kα radiation in the Bragg–Brentano ($\theta$–$2\theta$) geometry, and by micro-Raman spectroscopy using the 514.5 nm excitation line of an Ar$^+$ laser. The full-width at half-maximum (FWHM) of the diffraction lines were calculated by fitting with a pseudo-Voigt function while the Raman shifts were obtained by fitting the Raman bands with a Lorentzian function. The morphology and microstructure of the films were analysed by scanning electron microscopy (SEM) using an acceleration voltage of 20 kV, and their thickness measured with a Dektak stylus profilometer. Particle size distributions were calculated by digital analysis of the SEM micrographs recorded from the surface of the films.

## 3. Results and discussion

Because the substrate material is transparent to the KrF laser radiation and a low $Cr(CO)_6$ partial pressure was used, the deposition can be classified as a nearly pure photolytic surface mediated process. Only at the substrate level was some scattering of the laser radiation observed. As will be shown, the deposition of $Cr_2O_3$ was achieved within a very narrow range of the experimental parameters, although reproducibility was excellent. Best results were obtained for $F$=120 mJ cm$^{-2}$ and $R$=1.0. Films deposited using these experimental conditions are greenish, well adherent to the substrate and well crystallized, their thickness (*th*) following the laser beam transversal Gaussian profile, as can be seen in figure 1. Films deposited at $60 \leq F < 120$ mJ cm$^{-2}$ present rather low deposition rates and do not completely cover the irradiated region of the substrate; for $F<60$ mJ cm$^{-2}$ and/or $R>2.9$ no deposition was observed. On the other hand, when using laser fluences higher than 125 mJ cm$^{-2}$ substrate damage and/or film ablation occur, no matter the $R$ values employed.

The micro-Raman spectra and the XRD patterns of the films deposited at $F$=120 mJ cm$^{-2}$ as a function of the oxygen to chromium hexacarbonyl partial pressure ratio are shown in figure 2. A set of bands matching the Raman active modes of $Cr_2O_3$ [26-28] can be seen in figure 2a. The diffractograms of the same films (fig. 2b) show that those deposited at $R$=0.3 and 1.0 are polycrystalline, exhibiting several well resolved diffraction lines that could only be indexed to the rhombohedral α-$Cr_2O_3$ phase (JCPDS file nº 38-1479). No traces of other chromium oxide secondary phases were observed. Mean crystallite sizes of 146 nm and 95 nm were estimated for the films grown at $R$=0.3 and 1.0, respectively, using the (104) reflection at $2\theta = 33.602º$ and Scherrer's equation [29]. In contrast, the intensities of the diffraction lines of the films grown at $R$=0.1 and 2.9 are very low, suggesting a smaller amount of deposited material i.e., lower deposition rate and/or the growth of a prominent amorphous phase. However, the analysis of the micro-Raman spectra of the films (fig. 2a) shows that the $A_{1g}$ band is similar for the three films grown with $R \geq 0.3$ and only for the film deposited at $R$=0.1 is this band broader and red shifted, which is usually associated with structural disorder [27]. Therefore, the predominant factor determining the observed low intensity of the diffraction patterns for the films grown at $R$=0.1 and 2.9 should be the lower growth rate. Figure 3 shows the dependence of film thickness on the $R$ parameter of the films grown at $F$=120 mJ cm$^{-2}$. The thickness increases with $R$ up to $R$~1.5, as estimated by a 2nd order polynomial fit of the experimental data, decreasing sharply for higher pressure ratios. Films grown at $R$ values of 0.1 and 2.9 have similar thicknesses. Films deposited with $R$=1.0 are those with the highest measured deposition rate, reaching values of 0.1 nm s$^{-1}$, one order of magnitude higher than the deposition rates reported for $Cr_2O_3$ films deposited by PECVD at 300 ºC, using the same Cr precursor [13].

The influence of $R$ on the deposition trend may be understood taking into account the interactions between the reactive species on the surface of the substrate. Parnis *et al*. [30]

reported that the reaction rate of Cr with $O_2$, where Cr results from the photodissociation of $Cr(CO)_6$, increases linearly as a function of $p_{O2}$. Thus, a linear increase of the chromium oxide deposition rate with increasing $R$ would be expected if the interactions of the substrate surface with the gaseous atmosphere were irrelevant. In fact, deposition is initiated by one-photon dissociation of the hexacarbonyl to $Cr(CO)_x$ photoproducts which are adsorbed on the substrate surface within the laser beam spot and further dissociate to Cr [31]. When $O_2$ is present, the final reaction step is the surface reaction to form the chromium oxides. Although in the molecular state, $O_2$ is also competing for the same adsorption sites as $Cr(CO)_x$. The decrease in the available surface adsorption sites for the Cr precursor with increasing oxygen pressure leads to the decrease of the chromium oxide deposition rate. Therefore, the interplay of these two opposite effects explains the non-linear behaviour of the $Cr_2O_3$ deposition rate, in particular the existence of a maximum on film thickness as a function of $R$. On the other hand, low oxygen partial pressure regimes favour the growth of oxygen deficient films with higher structural disorder. This explains the broadening and red shift of the $A_{1g}$ Raman peak observed for the sample deposited at $R=0.1$, compared to the film grown at $R=2.9$, despite their similar thicknesses.

The effect of decreasing laser fluence on the growth of $Cr_2O_3$ films was studied by conducting new deposition experiments with different $R$'s using laser fluences of 80 mJ cm$^{-2}$ and 60 mJ cm$^{-2}$. The very low intensity of the diffraction lines in the XRD patterns (not shown), and the indexing of $Al_2O_3$ peaks (substrate material) in the micro-Raman spectra (fig. 4) of the films deposited with these fluences are a clear indication that the deposition rate decreases significantly with decreasing laser fluence. Actually, it was not possible to determine reliable thickness values for these films since often they do not fully cover the irradiated zone of the substrate. In particular for $F=60$ mJ cm$^{-2}$ and $R=2.9$ no deposition was observed. Regarding phase composition, the micro-Raman spectra presented in figure 4 show that the laser fluence has no significant effect on the material synthesised as no other phases besides $Cr_2O_3$ were identified.

Figure 5 shows the FWHM of the $Cr_2O_3$ $A_{1g}$ Raman shift as a function of $R$ for three different laser fluences. The films deposited at $F=120$ mJ cm$^{-2}$ and $R=1.0$ yield the narrowest $A_{1g}$ peaks and thus have the highest degree of crystallinity. Given the results described above, we turned our attention to the microstructure of the films deposited at laser fluence of 120 mJ cm$^{-2}$. Particle size and particle size distribution are important microstructural attributes of thin films since they can strongly affect the mechanical and tribological, chemical, optical and physical properties. The microstructure of the as-grown $Cr_2O_3$ films was studied by SEM in bright field mode. The films consist of a high density of irregular shaped particles, as can be seen from the micrographs of figure 6 for different $R$ values. In order to evaluate the mean particle size and to build the size distribution histograms of the as-deposited films, the lengths of about 400 particles were measured. Particle size distributions were obtained using the cumulative percentage method [32]. The histograms as well as their cumulative percentage for each $R$ value are also presented in figure 6. The particle size of the samples grown at $R$=2.9 and 1.0 follow a lognormal distribution with a mean particle size of 1.76 μm ($\sigma$=0.29 μm) and 1.85 μm ($\sigma$=0.24 μm), respectively, where $\sigma$ is the standard deviation. The latter result shows that for the film grown at $R$=1.0, particles are composed of about twenty nanocrystallites, according to the size of the coherent diffracting domains calculated from the XRD pattern of this sample. Furthermore, the lognormal distribution obtained for these two films strongly supports that they grow basically via a particle diffusion mechanism and subsequent coalescence [33]. Different particle distribution histograms were found for the samples deposited at lower oxygen partial pressure. Films deposited with $R$=0.3 show a rather flat particle size distribution and thus a high microstructure heterogeneity. A mean particle size of 2.02 μm ($\sigma$=0.74 μm) was estimated for this sample. On the other hand, films deposited with $R$=0.1 show a bimodal particle size distribution which can be attributed to the coexistence of crystallized particles with amorphous regions and/or with not fully crystallized aggregates, in agreement with the higher structural disorder observed for the films grown at low oxygen partial pressure, as previously described. In this case an average particle size of 1.70 μm ($\sigma$=0.36 μm) for the smaller particles and 2.72 μm ($\sigma$=0.12 μm) for the larger ones were obtained. Altogether, these results show that particle size and particle size distribution are strongly dependent on the reactant atmosphere composition used to grow the $Cr_2O_3$ single phase layers. For the majority of thin film applications it is crucial to have a particle size as homogeneous as possible and thus a narrow particle size distribution. This is indeed the case with the samples grown with $R$ values of 2.9 and 1.0, in particular the former, for which a smaller particle size standard deviation was found.

### 3.3 Conclusions

We have studied the synthesis of $Cr_2O_3$ thin films produced at room temperature by low pressure LCVD, using KrF laser radiation to promote the photodissociation of $Cr(CO)_6$ in dynamic atmospheres containing $O_2$. Although deposition was achieved within a very narrow range of the experimental parameters, the ability of this technique to achieve selective area growth of highly pure polycrystalline chromia was established. It was shown that the deposition rate and the degree of crystallinity of the as-deposited material increase with increasing laser fluence. On the other hand, it was shown that oxygen is essential for the growth of fully crystalline $Cr_2O_3$, despite appearing to compete with the $Cr(CO)_6$ photoproducts for the surface adsorption sites and thus there is a need for a fine balance between the partial pressures of both reactants. It was also shown that the film particle size distribution strongly depends on the $O_2$ to $Cr(CO)_6$ pressure ratio, low oxygen partial pressures inducing larger mean particle size and wider particle size distribution. Best results were obtained for $F$=120 mJ cm$^{-2}$ and $R$=1.0. Films deposited using these experimental parameters are polycrystalline with a narrow lognormal particle size distribution, which reflects their microstructural homogeneity. A deposition rate of 0.1 nm s$^{-1}$ and mean particles size of 1.85 μm with standard deviation of 0.24 μm were measured in such films.

**Acknowledgements**

The authors gratefully acknowledge the financial support of Fundação para a Ciência e Tecnologia (FCT) under contract POCTI/CTM/41413/2001 and Pluriannual contract with ICEMS. P.M. Sousa also acknowledges FCT for a PhD grant (BD16567/2004).

**References**

1. B. Kubota, J. Am. Ceram. Soc. 44 (1960) 239

2. J.L. Zang, J. Huang and C. Ding, J. Therm. Spray Technol. 7 (1998) 242.

3. E. Celik, C. Tekmen, I. Ozdemir, H. Cetinel, Y. Karakas, S.C. Okumus, Surf. Coat. Technol. 174/175 (2003) 1074.

4. T. Yu, Z.X. Shen, J. He, W. X. Sun, S. H. Tang, J.Y. Lin, J. Appl. Phys. 93 (2003) 3951.

5. S.A. Makhlouf, J. Magn. Magn. Mater. 272/276 (2004) 1530.

6. U. Balachandran, R.W. Siegel, Y.X. Liao, T.R. Askew, NanoStruct. Mater. 5 (1995) 505.

7. P. Kofstad, K.P. Lillerud, J. Electrochem. Soc. 127 (1980) 2410.

8. T. Maruyama, H. Akagi, J. Electrochem. Soc. 143 (1996) 1955.

9. R.S. Boorse, J.M. Burlitch, Chem. Mater. 6 (1994) 1509.

10. T. Maruyama, H. Akagi, J. Electrochem. Soc. 143 (1996) 1955.

11. F. Barradas-Olmos, J.R. Vargas-Garcia, J.J. Cruz-Rivera, Mater. Sci. Forum 386 (2002) 353.

12. F.K. Perkins, C. Hwang, M. Onellion, Y.G. Kim, P.A. Dowben, Thin Solid Films 198 (1991) 317.

13. J. Wang, A. Gupta, T.M. Klein, Thin Solid Films 516 (2008) 7366.

14. G. Gheorghies, L. Gheorghies, J. Optoelectron. Adv. Mater. 3 (2001) 107.

15. Y. Ito, K. Kushida, H. Takeuchi, J. Cryst. Growth 112 (1991) 427.

16. S. Noguchi, M. Mizuhashi, Thin Solid Films 77 (1981) 99.

17. X. Pang, K. Gao, F. Luo, H. Yang, L. Qiao, Y. Wang, A.A. Volinsky, Thin Solid Films 516 (2008) 4685.

18. S. Sasaki, Y.F. Zhang, O. Yanagisawa, and M. Izumi, J. Magn. Magn. Mater. **310** (2007) 1008.

19. P. Borisov, A. Hochstral, V.V. Shvartsman, W. Kleemann, P.M. Hauck, Integr. Ferroelectr. 99 (2008) 69.

20. M.A. Henderson, S.A. Chambers, Surf. Sci. 449 (2000) 135.

21. H. Mändar, T. Uustare, J. Aarik, A. Tarre, A. Rosental, Thin Solid Films 515 (2007) 4570.

22. R. Cheng, C.N. Borca, P.A. Dowben, S. Stadler, Y.U. Idzerda, Appl. Phys. Lett. 78 (2001) 521.

23. O. Conde, A.J. Silvestre, Appl. Phys. A 79 (2004) 489.

24. P.M. Sousa, A.J. Silvestre, N. Popovici, O. Conde, Appl. Surf. Sci. 247 (2005) 423.

25. M. M. Windsor, A.A. Blanchard, J. Am. Chem. Soc. 56 (1934) 823.

26. J. Mougin, T.L. Bihan and G. Lucazeau, J. Phys. Chem. Solids 62 (2001) 553.

27. J. Zuo, C. Xu, B. Hou, C. Wang, Y. Xie, Y. Qian, J. Raman Spectrosc. 27 (1996) 921.

28. S.I. Dolgaev, N.A. Kirichenko, G.A. Shafeev, Appl. Surf. Sci. 138/139 (1999) 449.

29. B.D. Cullity, S.R. Stock, Elements of X-ray diffraction, 3rd edition, Prentice-Hall, Inc., New Jersey, 2001, p. 170.

30. J.M. Parnis, S.A. Mitchell, P.A. Hackett, J. Phys. Chem. 94 (1990) 8152.

31. G.W. Tyndall, R.L. Jackson, J. Am. Chem. Soc. 109 (1987) 582.

32. M. Vopsaroiu, G.V. Fernandez, M.J. Thwaites, J. Anguita, P.J. Grundy, K.O.'Grady, J. Phys. D: Appl. Phys. 38 (2005) 490.

33. J. Soderlund, L.B. Kiss, G. A. Niklasson, C.G. Granqvist, Phys. Rev. Lett. 80 (1998) 2386.

**Figure captions**

**Figure 1** - Thickness profile of a film deposited with $F$=120 mJ cm$^{-2}$ and $R$=1.0.

**Figure 2** - Top panel: micro-Raman spectra of films deposited with $F$=120 mJ cm$^{-2}$ and different $R$ values. The $A_{1g}$ and $E_g$ vibration modes of chromia are assigned. S refers to the vibration modes of the substrate. a) $R$=2.9, $th$=0.6 μm; b) $R$=1.0, $th$=1.5 μm; c) $R$=0.3, $th$=1.0 μm; d) $R$ = 0.1, $th$=0.7 μm. Bottom panel: XRD patterns of the same films with the $Cr_2O_3$ diffraction planes assigned according to the JCPDS file nº 38-1479.

**Figure 3** - Film thickness as a function of $R$ for $F$=120 mJ cm$^{-2}$; (---) fitting curve with a 2nd order polynomial.

**Figure 4** - Micro-Raman spectra of films deposited using a) $F$=80 mJ cm$^{-2}$ and b) $F$=60 mJ cm$^{-2}$, at different $R$ values (bold face figures). The $A_{1g}$ and $E_g$ vibration modes of chromia are assigned. S refers to the vibration modes of the substrate.

**Figure 5** - Full-width at half-maximum of the $Cr_2O_3$ $A_{1g}$ Raman band *vs.* $R$ values for films deposited at different laser fluences.

**Figure 6** - SEM micrographs of the surface of films deposited with $F$=120 mJ cm$^{-2}$ and a) $R$=2.9; b) $R$=1.0; c) $R$=0.3; d) $R$=0.1. The corresponding particle size distribution histograms are displayed in the right side of each SEM image. A lognormal distribution function was used to fit the data of films a) and b).

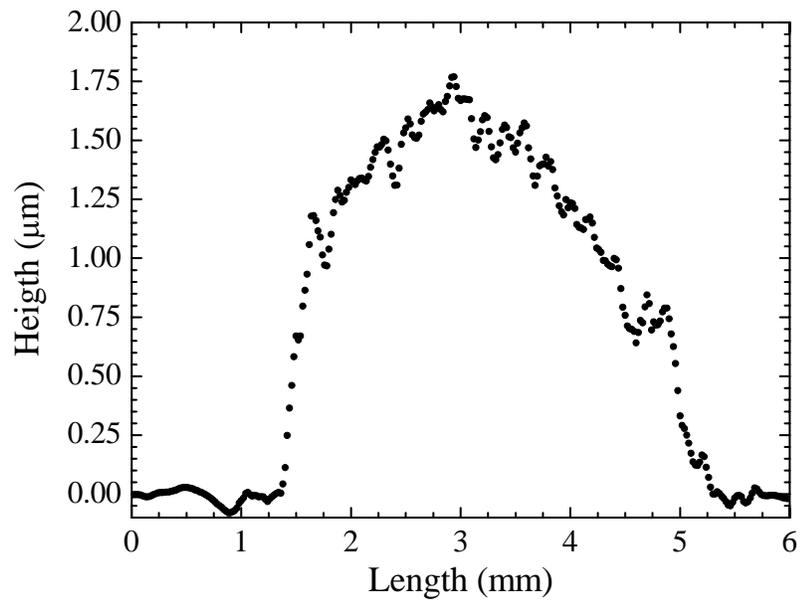

**Figure 1**

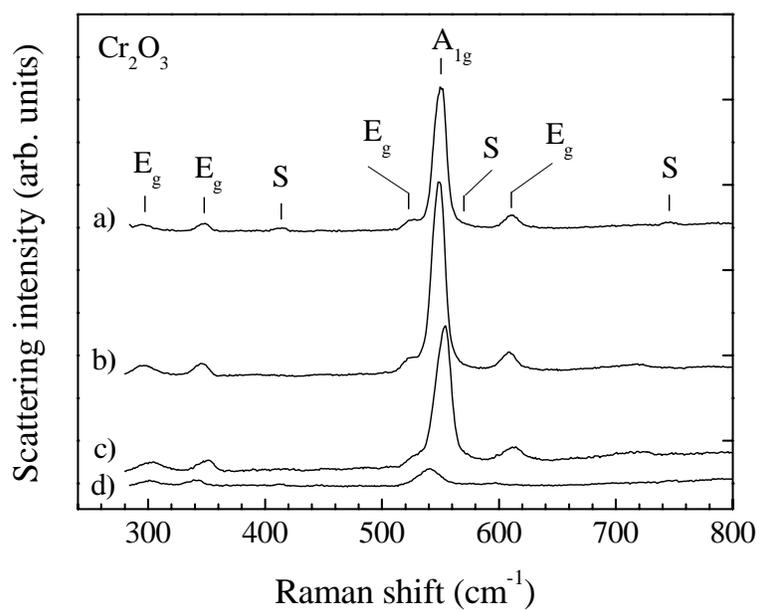

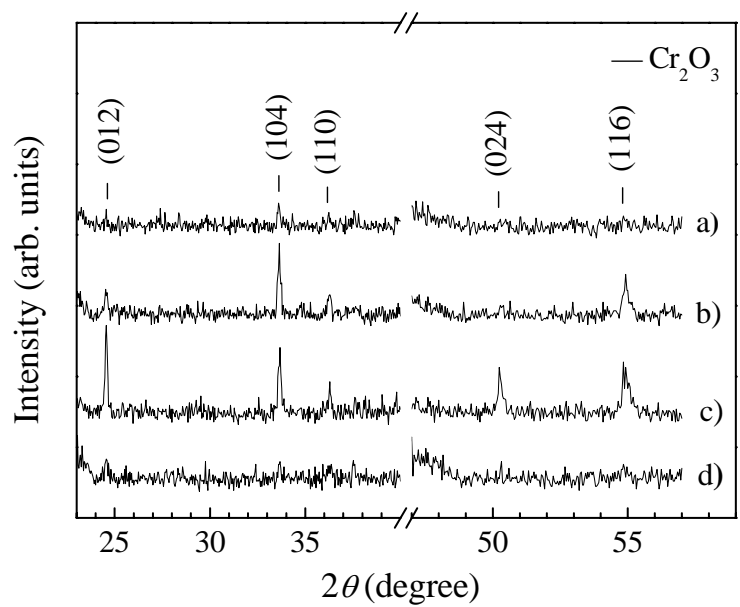

**Figure 2**

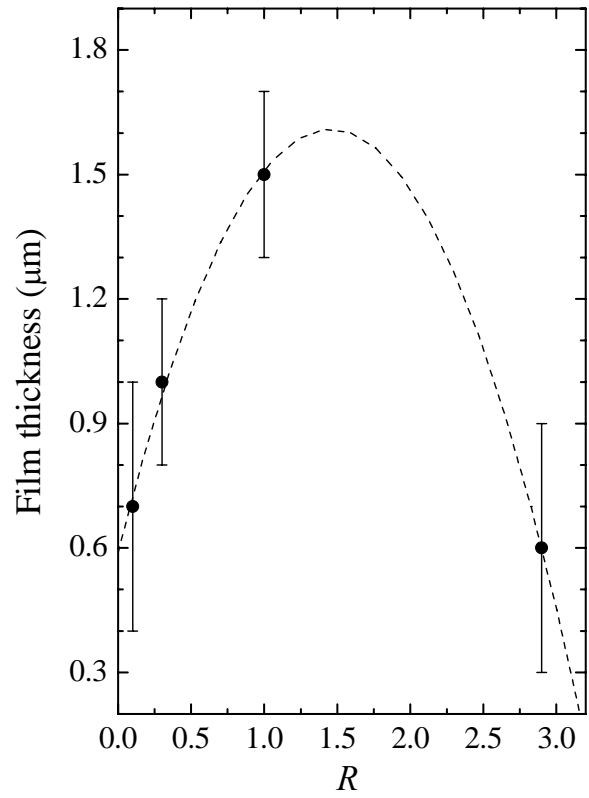

**Figure 3**

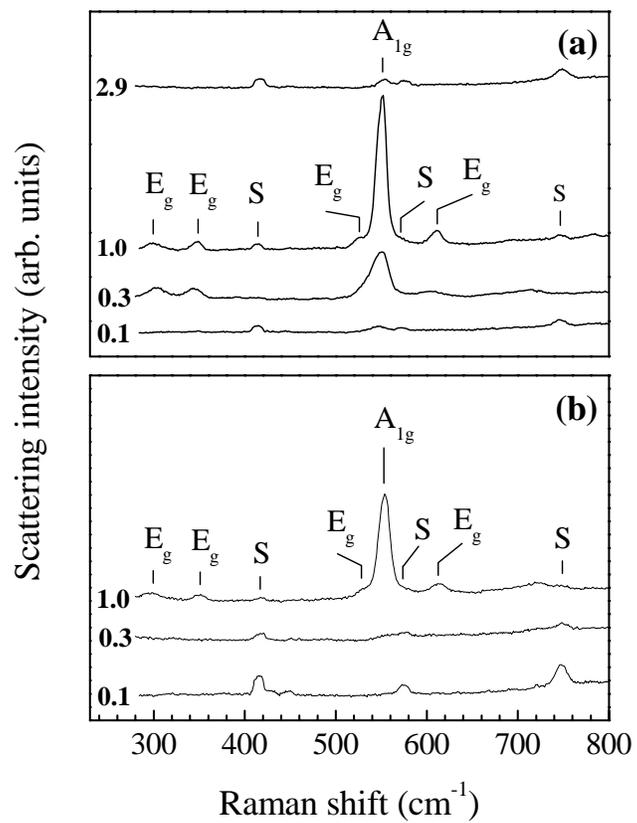

**Figure 4**

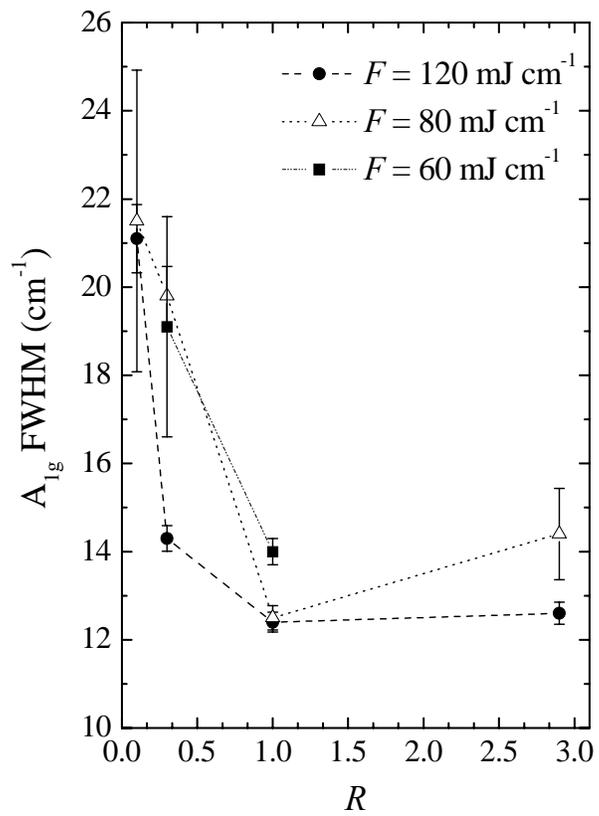

**Figure 5**

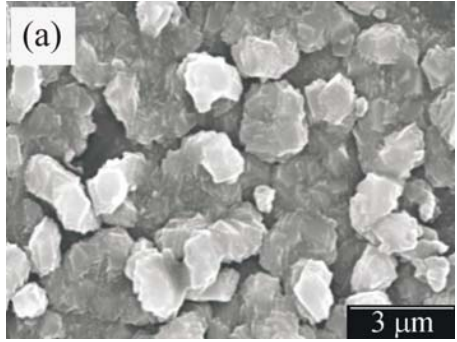
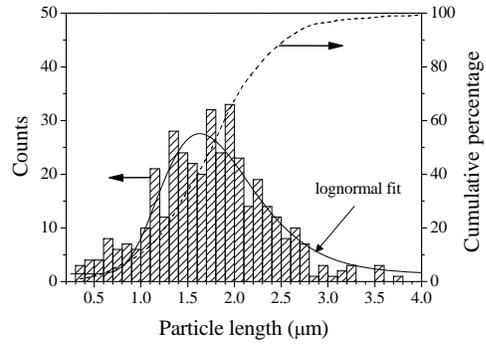
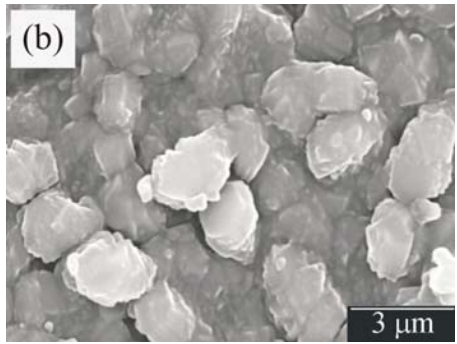
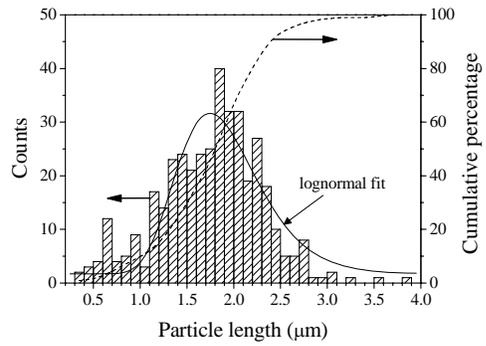
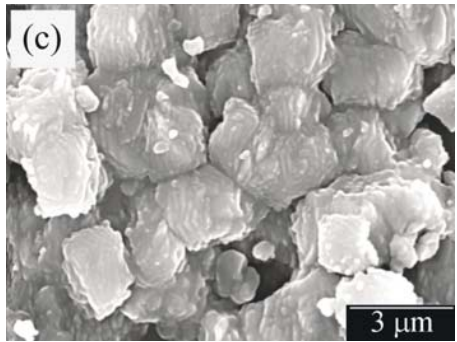
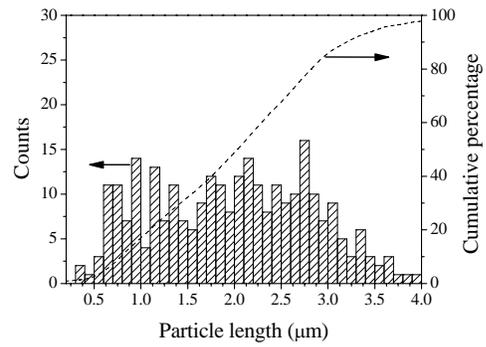
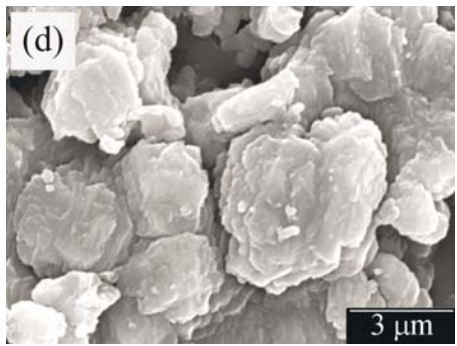
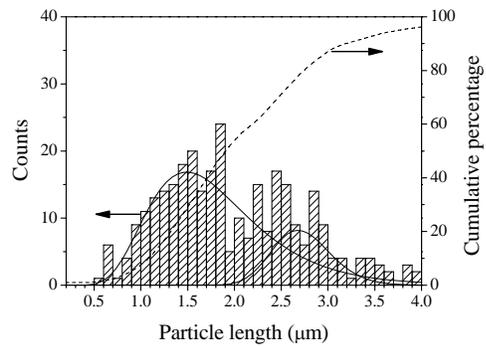

**Figure 6**